\documentclass{mn2e}
\usepackage{psfig}

\def\la{\mathrel{\hbox{\rlap{\hbox{\lower4pt\hbox{$\sim$}}}{\raise2pt\hbox{$<$}}}}}
\def\ga{\mathrel{\hbox{\rlap{\hbox{\lower4pt\hbox{$\sim$}}}{\raise2pt\hbox{$>$}}}}}
\def\Msun{\hbox{$\rm\thinspace M_{\odot}$}}
\def\Mjup{\hbox{$\rm\thinspace M_{\textrm{\tiny JUP}}$}}
\def\mwd{\hbox{$\rm\thinspace M_{\textrm{\tiny WD}}$}}

\def\tms{\hbox{$\rm\thinspace t_{\textrm{\tiny MS}}$}}

\begin{document}

\title[Imaging planets around nearby white dwarfs]{Imaging 
planets around nearby white dwarfs}

\author[M.\,R. Burleigh, F.\,J. Clarke and
S.\,T. Hodgkin]{M.\,R. Burleigh$^1$, F.\,J. Clarke$^2$ and
S.\,T. Hodgkin$^2$\\ 
$^1$ Department of Physics and Astronomy, University
of Leicester, University Rd., Leicester, LE1 7RH.\\ 
$^2$ Institute of Astronomy, Madingley Road, Cambridge, CB3 0HA.\\ 
email: mbu@star.le.ac.uk, fclarke,sth@ast.cam.ac.uk}


\maketitle

\begin{abstract}
We suggest that Jovian planets will survive the late
stages of stellar evolution, and that white dwarfs will retain planetary
systems in wide orbits ($\ga5$\,AU).  Utilising evolutionary models for
Jovian planets, we show that infra-red imaging with 8m class telescopes of
suitable nearby white dwarfs should allow us to resolve and detect
companions $\ga3\Mjup$.  Detection of massive planetary companions to
nearby white dwarfs would prove that such objects can survive the final
stages of stellar evolution, place constraints on the frequency of main
sequence stars with planetary systems dynamically similar to our own and
allow direct spectroscopic investigation of their composition and
structure.

 
\end{abstract}


\section{Introduction} 

Over 70 extra-solar planets have now been detected since the discovery of a
companion to the solar-type star 51~Peg in 1995 (Mayor \& Queloz 1995).
All of these planets have been discovered via the radial velocity
technique, in which the planet's presence is inferred by the central star's
motion around the system's barycentre. Instrumental and intrinsic noise
limit the sensitivity of these studies to Saturn-mass companions in
short-period orbits, with more massive companions detectable in more
distant orbits.  The duration of current programmes limits this to
$\sim$3.5\,AU (Fischer et~al. 2002), although some systems do display
interesting trends indicative of more distant companions.  Therefore, we
have little information on systems with massive companions at large radii,
such as our own Solar System. The only constraint on Jovian-like systems
comes from micro-lensing statistics, which suggest less than $1/3$ of M
stars have Jupiter-like planets orbiting at $3-7$\,AU (Albrow et~al.\
2001).



The planets discovered by the radial velocity technique are
not open to further {\em direct} study, due to their close proximity to
the much brighter parent star. The only exception to this is the
transiting companion to HD209458 ($P_{\rm orb}\approx3.5$~days). Transit
photometry shows this planet is a gas-giant, and Charbonneau et~al. (2002)
have recently detected sodium in its atmosphere. Still, the planet itself 
cannot be directly imaged.

Several groups have conducted imaging surveys of nearby main sequence stars
to search for low mass companions, including the use of adaptive optics,
coronographs and space-based observations with HST (e.g.~Turner et~al.\
2001, Neuh\"auser et~al.\ 2001, Oppenheimer et~al. 2001, Kuchner \& Brown
2000, Schroeder et~al.\ 2000). However, the extreme contrast
($\geq$\,20~magnitudes) and small separations (5AU$=1\arcsec$ at 5pc)
between main sequence stars and Jovian planets makes sensitive surveys very
difficult.  To date, no planetary mass companions to nearby stars have been
imaged, although several brown dwarf companions have been detected via
direct imaging (e.g. Nakajima et~al.\ 1995).

The end state of main sequence stars with M$\leq$8\Msun, white dwarfs, are
typically 10$^3$--10$^4$ times less luminous than their progenitors. Thus,
there is potentially a strong gain in the brightness contrast between a
planet and a white dwarf when compared to a main sequence star, assuming
that planets can survive the late stages of stellar evolution. The gain in
contrast is strongest in the mid-infrared, where the planet's thermal
emission peaks well into the white dwarf's Rayleigh-Jeans tail.

Indeed, Ignace (2001) has suggested that excess infra-red emission could be
detectable from a hot Jupiter orbiting a 10,000K white dwarf at a distance
of $\sim10^3$ white dwarf radii and with an orbital period of
$\sim10$~days. Chu et~al. (2001) have also suggested that, near a hot
UV-bright white dwarf, the atmosphere of a Jovian planet would be
photoionized and emit variable hydrogen recombination lines, which may be
detected by high-dispersion spectroscopic observations.
However, both these methods rely on the planet being in a close (0.01--2AU)
proximity to the white dwarf, where it would be difficult to resolve.

In this letter, we discuss the potential for imaging planetary companions
in wide ($>$5AU) orbits around nearby white dwarfs.  In section 2 we
investigate the probability of a planetary system surviving the late stages
of stellar evolution. We discuss the detectability of any surviving planets
in section 3, suggest suitable target white dwarfs in section 4, and
discuss the expected frequency of wide planetary companions to white dwarfs
in section 5.

\section{Planetary systems in the post-main sequence phase}

Any planetary companion to a white dwarf must have survived the red giant
branch (RGB) and asymptotic giant branch (AGB) phases of stellar
evolution. During these stages, the white dwarf progenitor swells up to a
few hundred solar radii in size and undergoes significant mass loss.  What
happens to a planet during this time depends on the initial orbital
separation, the stellar mass loss rate, the total mass lost, tidal forces,
and interaction with the ejected material.

Any planet in an initial orbit within the final extent of the red giant's
envelope will be engulfed and migrate inwards.  Livio \& Soker (1984)
showed that these planets will either completely evaporate or accrete mass
and become a close companion to the eventual white dwarf.

Planets in wider orbits that avoid direct contact with the expanding red
giant envelope (such as the Jovian planets of our own solar system) will
have a greater chance of survival, migrating outward as mass is lost from
the central star. However, stars more massive than the Sun lose more than
half of their original mass in evolving to the white dwarf stage (Weidemann
1987, Wood 1992). If over half the mass of the central star is lost {\em
instantaneously} then any accompanying planets will likely escape, since
their orbital velocity will suddenly exceed the escape velocity of the
system. Thus, we might naively expect no white dwarfs to retain planetary
systems.

Of course, in reality, main sequence stars of M$\le$8.0\Msun\ do not evolve
into white dwarfs instantaneously. Mass loss on the RGB and AGB, and final
ejection of the planetary nebula is a relatively slow process.  The Sun,
for example, will lose around $0.2\Msun$ on the RGB in $\sim$\,10$^6$ years,
and up to another $0.2\Msun$ on the AGB on a timescale of up to $\sim$\,10$^5$
years. The final thermal pulse, in which the cool red giant develops a
super-wind and finally ejects the remainder of its
envelope, exposing the core, is poorly understood but probably occurs on a
timescale of a few $\times$\,10$^4$~years (Soker 1994).

The dynamical timescale for a planet to react to the change in the mass 
of the central star, and thus the gravitational force between the star and 
planet, is given by;

\begin{equation}
t_{dyn} \simeq \frac{a}{v_{orb}}
\label{eq:dyntimscl}
\end{equation}

\noindent
where $a$ is the semi-major axis of the orbit and 
$v_{orb}$ is 
the orbital velocity, and; 

\begin{equation}
v_{orb} = \sqrt{\frac{\textrm{G}(M_1 + M_2)}{a}}
\label{eq:vorb}
\end{equation}

\noindent
where $M_1$ is the mass of the central star and 
$M_2$ is the mass of the planet. 
Assuming $M_1 >> M_2$; 

\begin{equation}
t_{dyn} \simeq 14{\textrm{yr}}
\left(\frac{a}{20{\textrm{AU}}}\right)^{3/2}
\left(\frac{M_1}{{\textrm{M}}_\odot}\right)^{-1/2}
\label{eq:dyntimscl2}
\end{equation}

This is much less than the mass loss timescales mentioned above. Even if a
$2\Msun$ star lost its entire envelope ($1.4\Msun$) during a short PN phase
of $10^3$ years (which it doesn't, only the last few tenths of a solar
mass), the orbit would not become unbound. Thus, to a first approximation,
the orbits of planets which do not interact directly with the red giant
will simply expand adiabatically by a maximum factor M$_{\rm MS}$/M$_{\rm
WD}$ (Jeans 1924, 
Zuckerman \& Becklin 1987). For example, for a main sequence star
$\sim$\,2\Msun\ and a white dwarf $\sim$\,0.6\Msun, the orbit would expand
by a maximum factor $\sim$\,3.  A planet in a Jupiter-like orbit would be
relocated to $\sim$\,15\,AU, and a planet in a Neptune-like orbit to
$\sim$\,90\,AU.

%

For planets in orbits $\leq$10\,AU, tidal forces between the planet and red
giant are important in slowing orbital migration by transfering angular
momentum from the orbit to stellar rotation. For example, Soker (1994)
shows that, neglecting tidal forces, as the Sun evolves to a white dwarf
Jupiter's orbital radius will expand from its present value of
$a=1118R_\odot$ to $a=1860R_\odot$. However, tidal forces on the RGB will
counteract this expansion, reducing this final orbital radius by $\sim3$\%,
and by a further $\sim14$\% on the upper AGB. For planets within 5\,AU,
Livio \& Soker (1983) showed that tidal interactions will actually cause
them to migrate {\em inwards}.  Soker (1996) suggests that tidal
interactions between evolving red giants and low mass companions could be
responsible for producing the large fraction of elliptical PN observed,
indicating low mass companions may be common amoung PN progenitors.

The orbits of distant planets will also be subjected to drag
resulting from their interaction with the material ejected by the central
star. Duncan \& Lissauer (1998) define the total amount of material
impacting a planet during post-main sequence evolution as;

\begin{eqnarray*}
M_{{\textrm{\small hit}}}({\textrm{M}}_\odot) & = & \frac{R_{p}^2}{12a^2}
\left[1 - \left(\frac{M_{\tiny WD}}{\Msun}\right)^3\right] \\ & & \times
\left(\frac{v_{sw}^2 + v_{orb}^2}{v_{sw}^2}\right)^{1/2} \left(1 +
\frac{v_{esc}^2}{v_{sw}^2 + v_{orb}^2}\right)\\
\label{eq:accrate}
\end{eqnarray*}

\noindent 
where $v_{sw}$, $v_{orb}$ and $v_{esc}$ are the velocities of the
stellar 
wind, the planet and the escape velocity of the planet respectively,
and $R_{p}$ is the radius of the planet's magnetosphere. 
Evaluation of this equation, however, shows that gas giants will collide
with significantly less than 1\% of their own mass over the entire red
giant phase. The orbits will decay inwards only slightly as a result.

Duncan \& Lissauer (1998) have made detailed simulations of the effects of
post-main sequence mass loss on the stability of the solar system and for
planetary systems around more massive stars. They show that the orbits of
the giant planets will probably be stable for tens of billions of years
subsequent to the Sun's death. However, planetary systems dynamically
similar to our own around somewhat more massive stars may eventually be
de-stabilised.  Duncan \& Lissauer's calculations show that giant planets
around, for instance, an initially $4\Msun$ star, will be liberated by
large-scale chaos in less than a Hubble time.

Finally, Soker (1999) points out that the newborn planetary nebula central
star is hot (T$_{\rm eff}\ga100,000$K) and a strong source of ionizing soft
x-ray and UV radiation.  The interaction of this radiation with a planet
may lead to ablation of the planetary atmosphere, but this will only be
effective for planets having an escape velocity $v_{esc} \la c_{s}$,
where $c_{s} =15$\,km\,s$^{-1}$ 
is the sound speed of the ionised planetary material. For
a Jupiter-mass planet $v_{esc}=61$\,km\,s$^{-1}$ and
thus there will be negligible ablation.

It seems likely, then, that giant planets in initially distant (a$>$5AU)
orbits around main sequence stars $1-8\Msun$ will survive the late stages
of stellar evolution, and remain in orbit around the remnant white dwarfs
at increased orbital radii for at least several billion years.

\section{Detecting planets around nearby white dwarfs}

Our ability to detect an extra-solar planet around a nearby white dwarf
will depend on its intrinsic luminosity, which in turn depends on its age,
its distance from us, and its separation from the parent star.

\subsection{Luminosities of extra-solar planets}

Burrows et~al.\ (1997) have made nongray calculations of the expected
spectra, colours and evolution of solar-metallicity massive planets. We
have used these models to predict the infra-red magnitudes of planets
around nearby white dwarfs.
The Burrows et~al.\ models describe the evolution of a planet in isolation,
and do not include the effects of thermal insolation by, or reflected light
from, a parent star. In reality, gas-giant planets will have been
kept warm throughout their lives  
by the white dwarf progenitors. They will also have
accreted matter during the RGB/AGB/PN phases. Both effects would act to
increase their temperatures and luminosities, although 
in the white dwarf phase reflected light 
makes a negligible contribution to the planet's brightness. 
Our predicted luminosities are, therefore, lower limits.






\subsection{The ages of white dwarfs}

The white dwarf cooling age and progenitor main sequence lifetime are the
dominant timescales determining the brightness of any planetary companions,
as they are significantly longer than the RGB or pre-main sequence
phases.

White dwarf cooling ages are dependent on a number of parameters including:
effective temperature, mass (usually derived from the surface gravity),
composition of the core and the thickness of the surface H- (or He-)
dominated atmosphere. Temperature, surface gravity and atmospheric
composition are usually determined through fitting model atmospheres to
optical and UV spectra (e.g.~Bergeron, Saffer \& Liebert 1992, Marsh
et~al. 1997, Finley, Koester \& Basri 1997). The mass is then inferred from
the surface gravity measurement using evolutionary models (e.g.~Wood 1992,
1995). These same models also give the cooling age. The cooling ages and
masses of many nearby white dwarfs have been calculated in this manner by
Bergeron, Leggett \& Ruiz (2001, BLR).  For nearby white dwarfs not
included in the BLR sample (generally, those hotter than $\sim10,000$\,K),
we have estimated cooling ages from published temperature and mass
measurements.

The relationship between the mass of a white dwarf and its main sequence
progenitor, the initial-final mass relation, has been investigated by a
variety of authors. We have adopted model A from Wood (1992):

\begin{equation}
{\textrm{M}}_{\textrm{\tiny MS}} = 10.4\ln\left[\frac{\left(\textrm{M}_{\textrm{\tiny WD}}/\Msun\right)}{0.49}\right] \Msun
\label{eq:massrelation}
\end{equation}

\noindent
Wood (1992) also gives a simple formula for
estimating main sequence lifetimes:

\begin{equation}
{\rm t}_{\textrm{\tiny MS}} = 10\left(\frac{\textrm{M}_{\textrm{\tiny MS}}}{\Msun}\right)^{-2.5} {\textrm{Gyr}}
\label{eq:mslifetime}
\end{equation}

\noindent
The white dwarf cooling age and the main sequence lifetime can then be
combined to give the {\em system} age of local white dwarfs. We use this
age, combined with the distance to the white dwarf, to estimate the
magnitude of planets.




\section{Suitable targets among nearby white dwarfs}

There are 118 white dwarfs currently catalogued within 20pc of the Sun
(Holberg et~al. 2001). All of these systems are close enough to allow us to
resolve planetary companions in $\sim$\,5--100\,AU orbits from the ground.
For example, at 10\,pc an orbit of 10\,AU is equivalent to a maximum
separation of $1\arcsec$ on the sky.  Planets in wider orbits will be even
easier to resolve.  Adaptive optics, coronographs and techniques such as
nulling interferometry will increase the chances of resolving closer
companions.

\begin{figure}
\caption{The relationship between white dwarf mass and main-sequence
lifetime (from equations~\ref{eq:massrelation} and
\ref{eq:mslifetime}). The steepness of the relationship in the range
0.55$<$\mwd$<$0.6 shows why we should not rule out white dwarfs with mass
estimates in this range, despite the indications their progenitors had a
long main-sequence lifetime.}
\label{fig:mass-mstime}
\psfig{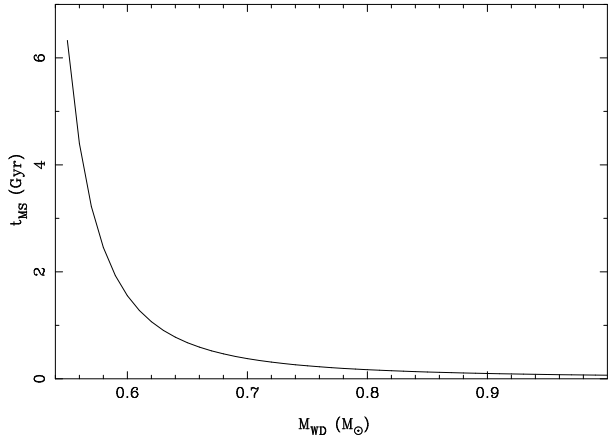}

\end{figure}

However, only those white dwarfs 
young enough for Jovian companions to still be
detectable will make suitable targets for extra-solar planet searches.  We
are most likely to be able to detect an extra-solar planet around a
relatively warm and young ($<$a few\,$\times10^8$\,yrs), massive
($>$0.6\Msun) white dwarf descended from a relatively massive progenitor
($>$2\Msun), since its main sequence lifetime is short ($<$1.8\,Gyr).  For
\mwd$<$0.6\Msun, the main sequence lifetime is very sensitive to errors in
the \mwd\ estimate (see Figure~\ref{fig:mass-mstime}). Therefore, we would
be unwise to rule out white dwarfs with mass estimates below 0.6\Msun, as
they may be descended from more massive stars than we expect.

Table~\ref{tab:wdtable} lists the top 10 candidates among nearby white
dwarfs for a search for planetary companions, considering the main sequence
lifetime \tms, the white dwarf cooling age t$_{\textrm{\tiny COOL}}$ and
the distance.  Also included in the table are the estimated J band
magnitude of 3, 5, \& 10\Mjup\ planets in each system.

\begin{figure}
\caption{Magnitudes of a 5\Mjup\ planet around nearby white dwarfs
(Table~\ref{tab:wdtable}) in the near-IR J,H,K,L$^\prime$ and M bands. Also
plotted are the nominal 
limiting sensitivities (1hr, 5-$\sigma$) for Gemini-N+NIRI
(solid lines) and VLT+ISAAC (dashed lines).}
\label{fig:magsroundwds}
\psfig{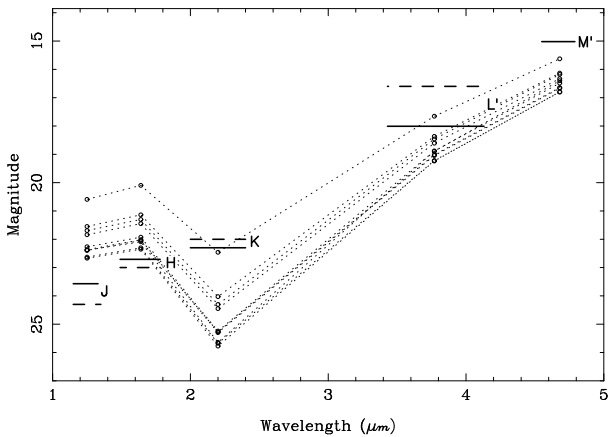}
\end{figure}

In Figure~\ref{fig:magsroundwds} we plot the magnitude of a 5\Mjup\ planet
around these white dwarfs in several near-IR bands (J,H,K,L$^\prime$ \&
M). Also plotted are the nominal sensitivity limits for current near-infrared
imaging instruments (from on-line exposure time estimates); 
Gemini-North $+$ NIRI (solid lines) and VLT $+$ ISAAC
(dashed lines). We can see that although the planets are brighter in the M
band (5$\mu$m), the instruments are only sensitive enough in the J and H
bands (1.1 \& 1.6$\mu$m).

\begin{table*}
\caption{Top 10 candidate white dwarfs for detecting
Jovian planets. White dwarf parameters are taken from; Bergeron, Saffer
\& Liebert (1992; BSL92); Wickramasinghe \& Ferrario (2000; WF00); Bergeron 
et~al. (1995; B95); Bergeron, Leggett
\& Ruiz (2001; BLR01); Bragaglia, Renzini \& Bergeron (1995; BRB95) and 
McMahan (1989; M89).}
\label{tab:wdtable}
\begin{tabular}{lccccccccccc}
\hline
WD & Distance & \mwd & t$_{\textrm{\tiny COOL}}$ & $M_{\rm MS}$ & \tms 
& t$_{\textrm{\tiny TOTAL}}$ & \multicolumn{3}{c}{Planet Mag (J)} & Ref. \\
	& (pc) & (\Msun) & (Gyr) & (\Msun) & (Gyr) & (Gyr) 
& 3\Mjup & 5\Mjup & 10\Mjup & \\
\hline
\hline
WD1134$+$300 & 15.32 & 0.90 & 0.15 & 6.3 & 0.1 & 0.25 & 22.7 & 20.6 & 18.5 & BSL92\\
WD1900$+$705 & 12.98 & 1.00 & 0.5 & 7.4 & 0.1 & 0.6 & 23.8 & 21.6 & 19.9 & WF00\\
WD1647$+$591 & 10.97 & 0.69 & 0.3 & 3.6 & 0.4 & 0.7 & 24.0 & 21.7 & 20.2 & B95 \\
WD0644$+$375 & 15.41 & 0.66 & 0.1 & 3.1 & 0.6 & 0.7 & 24.2 & 21.8 & 20.3 & BLR01\\
WD0310$-$688 & 10.15 & 0.63 & 0.2 & 2.6 & 0.9 & 1.1 & 24.7 & 22.3 & 21.2 & BRB95\\
WD2047$+$371 & 18.16 & 0.65 & 0.3 & 2.9 & 0.7 & 1.0 & 24.8 & 22.4 & 21.2 & M89 \\
WD2105$-$820 & 13.52 & 0.75 & 0.8 & 4.4 & 0.2 & 1.0 & 24.8 & 22.4 & 21.2 & BLR01\\
WD1142$-$645 & 11.19 & 0.75 & 1.4 & 4.4 & 0.2 & 1.6 & 24.9 & 22.4 & 21.6 & BLR01\\
WD2007$-$219 & 18.22 & 0.69 & 0.8 & 3.6 & 0.4 & 1.2 & 25.1 & 22.6 & 21.6 & BRB95 \\
WD1236$-$495 & 16.39 & 1.0 & 1.2 & 7.4 & 0.1 & 1.3 & 25.1 & 22.7 & 21.7 & BLR01\\
\hline
\end{tabular}
\end{table*}

\section{The expected frequency of planets in wide orbits around 
white dwarfs}

Assuming that massive planets born in initially wide ($\ga5$\,AU) orbits
remain bound to white dwarfs, then the expected frequency of massive
planets around nearby white dwarfs is simply a function of the frequency of
solar-like and more massive main sequence stars with Jovian-like planetary
systems.  Unfortunately, we currently have little observational information
on these systems, although it is possible to extrapolate from existing data
to make some estimate of what we might expect. Lineweaver \& Grether
(2002, LG) have analysed the statistics of planets detected via radial
velocity techniques, and estimated the number of Jovian-like systems we
might expect to find. They have produced a fit to the completeness-correct
period distribution of planets so far detected, in the mass range
$0.84\Mjup<M\sin i<$13\Mjup;

\begin{equation}
\frac{{\textrm{d}}N}{{\textrm{d}} \log{P}} = a\log{P} + b
\label{eq:dndlogp}
\end{equation}

\noindent
where $a = 12\pm3$ and $b = -9\pm2$. The initial orbital periods 
of interest range from $\sim$\,10\,yr, to avoid direct interaction with
the expanding red giant envelope, to $\sim$\,200\,yr, where we might expect
Jovian planet formation to end. Integrating equation~\ref{eq:dndlogp} over
this range, we estimate there to be $\sim$55$\pm$16 Jovian planets per
$\sim$1000 white dwarf systems. However, for a significant sample of nearby
white dwarfs we can only detect planets $\ga5\Mjup$ with current
technology.  LG fit the observed planetary mass function as;

\begin{equation}
\frac{{\textrm{d}}N}{{\textrm{d}}\log(M\sin i)} =  a\log(M\sin i) + b
\label{eq:dlogm}
\end{equation}

\noindent
where $a = -24\pm$4. To produce 55$\pm$16 planets over the mass range
$0.84\Mjup<M\sin i<13\Mjup$\, 
we need to set $b = -58.4\pm19.7$. Therefore,
from equation~\ref{eq:dlogm}, we expect $\sim$15$\pm$6 
planets $\ga5\Mjup$ 
per 1000 white dwarfs. Hence, we may expect $\sim$1--2\% of white
dwarfs to possess one or more planets $\ga5\Mjup$.  Note that we treat
$M\sin i$ as equivalent to $M$, so this predicted number density is a lower
limit.

This analysis is, however, based on statistics for planets in orbits
$\leq$\,3.5\,AU.  It is quite possible that laws derived for planets in
this area of orbital parameter space around $\sim$1\Msun\ stars do not
apply to Jovian planets at large radii ($>$5\,AU) around more massive stars
($2-8\Msun$).

\section{Discussion}

We have suggested that Jovian planets will survive the
late stages of stellar evolution, and that some nearby white dwarfs 
possess planetary systems in wide orbits. Utilising evolutionary models for
Jovian planets, we have shown that infra-red imaging of suitable nearby
white dwarfs should allow us to resolve and detect companions M$\ga$3\Mjup\
with 8m class telescopes. The best candidates for an observational search
are relatively hot (young), massive ($>0.6\Msun$) white dwarfs, as they
have the shortest overall system age.
However, given that we have neglected potential heating mechanisms such as
thermal insolation and accretion from the red giant's wind, Jovian planets
around nearby white dwarfs are likely to be brighter than we have suggested
in Figure~\ref{fig:magsroundwds}, and may be detectable around
$\sim0.55\Msun$ white dwarfs descended from progenitors as low in mass as
$\sim1.2\Msun$.

Our ``top ten'' list of targets among nearby white dwarfs (Table~1)
includes degenerates descended from main sequence progenitors from
2.6\Msun~$-$~7.4\Msun, corresponding to spectral types early A and B. The
existence of a dust disk around Vega (A0V) strongly suggests that planetary
systems may exist around such early-type stars, although they are not the
subject of current radial velocity searches.


Detection of massive planetary companions to nearby white dwarfs would
prove that such objects can survive the final stages of stellar evolution,
place constraints on the frequency of main sequence stars with planetary
systems dynamically similar to our own, and allow direct spectroscopic
investigation of their composition and structure.

An unsuccessful search would suggest one of, or a combination of, 
a number of scenarios, including: 
(a) that planets $\ga3\Mjup$ are fainter 
than expected from the
evolutionary models we have used; (b) that, at least for a sub-sample
of targets, any accompanying planets have too small a projected separation
from the white dwarf to be resolved; (c)
that massive Jupiters in wide orbits 
around 1--8\Msun\ main sequence stars are not common; 
(d) that these planets  
do not remain bound to white dwarfs. This last scenario
would raise the possibilty that the Galaxy contains a population of
isolated Jovian-mass planets ejected by white dwarfs during the final
stages of their evolution.

\section{Acknowledgements}

MRB, FJC and STH acknowledge the support of PPARC, UK. We would like to thank 
Eduardo Martin for an illuminating discussion, and James Murray and 
Andrew King for comments concerning orbital evolution.

\end{document}